\documentclass[aps,prl,twocolumn,groupedaddress,floatfix]{revtex4-1}
\usepackage{graphicx}
\usepackage{color}
\begin{document}

\title{Negative differential resistance in scanning tunneling microscopy: simulations 
on C$_{60}$-based molecular overlayers}
\author{Frederico D. Novaes$^1$}
\author{Manuel Cobian$^{1,2}$}
\author{Alberto Garc\'{\i}a$^1$}
\author{Pablo Ordej\'on$^3$}
\author{Hiromu Ueba$^4$}
\author{Nicol\'as Lorente$^{3}$}
 \affiliation{$^1$ Instituto de Ciencia de Materiales de Barcelona (CSIC), E-08193 Bellaterra, Spain\\ 
$^{2}$ LPMCN, UCBL, CNRS, UMR5586, 69622 Villeurbanne, France\\
$^{3}$ Centre d'Investigaci\'o en Nanoci\`encia i Nanotecnologia (CSIC-ICN), E-08193 Bellaterra, Spain\\
$^4$ Division of Nano Science, Graduate School of Science and
Engineering, University of Toyama, Toyama, Japan
}
\date{\today}

\begin{abstract}
We determine the conditions in which negative differential resistance
(NDR) appears in the C$_{60}$-based molecular device of  [Phys. Rev. Lett.
{\bf 100}, 036807 (2008)] by means of ab-initio electron-transport
simulations.  Our calculations grant access to bias-dependent intrinsic
 properties of the molecular device, such as electronic levels
and their partial widths. We show that these quantities depend on
the molecule-molecule and molecule-electrode interactions of the
device. Hence, NDR can be tuned by modifying the bias behavior of levels
and widths using both types of interactions.  
\end{abstract}

\maketitle

%Introduction
Since the creation of the first tunnel diode~\cite{Esaki}, a
non-linear $I$--$V$ characteristic, particularly showing negative
differential resistance (NDR) is the cornerstone in two-terminal
devices~\cite{Tsu,Sze}.  NDR permits the electronic current of these
devices to be modified by the applied bias in a complex way.  This
active control of the current  motivates a long search for NDR in
molecular devices~\cite{Aviram,Lyo,Bedrossian,Chen,Zeng,Rinkio,Zheng}
where the scanning tunneling microscope (STM) plays an important
role~\cite{Lyo,Bedrossian}.  The STM can vary applied bias and tunneling
current in an independent manner, yield atomically precise data and still
be a two-terminal device.  
More recently, STM studies of C$_{60}$ molecules have spurred a
lot of interest because of the NDR capabilities of C$_{60}$-based
devices~\cite{Zeng,Zheng,Grobis,Franke,Isabel}.  

As early as the first NDR cases were obtained in STM junctions, the
existence of sharp resonances on both electrodes were used to explain
the decrease of current as the bias increased~\cite{Lyo,Bedrossian,Xue}.
However, Grobis and co-workers~\cite{Grobis} explained their own NDR data
by uncovering yet another mechanism: the voltage-dependent increase of the
tunneling barrier height. Recently, one more mechanism has been advanced
in order to explain the dependence of NDR on the tip's material, namely
orbital matching between molecule and tip~\cite{Chen07}.  This last
mechanism has been thoroughly studied and validated, while showing
the need of sharp STM tips~\cite{Shi}.  However, in most molecular
devices, the accurate evaluation of the voltage drop across the device
is crucial. As a matter of fact, Tu and co-workers~\cite{Tu} showed
the paramount importance of the actual voltage-dependent barrier for
the presence or absence of NDR.  This spread of mechanisms shows that
many parameters change among different molecular junctions and it is
of fundamental importance to understand the key ingredients in NDR and
thus the final functionality of possible molecular devices.

In this letter, we study the molecular device of Ref.~\cite{Franke}
unraveling the mechanisms leading to the measured NDR~\cite{Franke}.
The simulated molecular device consists of
C$_{60}$ molecules partially decoupled from a Au (111) substrate by
spectator 1,3,5,7-tetraphenyladamantane (TPA) molecules.  
Despite the complexity of the simulations, we can extract the molecular
levels and their coupling with the electrodes while describing their bias
dependence. This allows us to have unprecedented insight in the actual
physics of a molecular device with a quantitative description. 
We show that we can tune the size of the NDR signal
by modulating the interplay of the bias dependence of the levels with the bias
dependence of their partial widths which ultimately depend on 
molecule-molecule and molecule-electrode interactions.

In the Landauer formalism~\cite{Landauer}, the electronic current, $I$,
is given in terms of the electron transmission function $T(E,V)$ with
$E$ the electron energy and $V$ the applied bias ($e$ and $h$ being the
electron charge and Planck's constant, respectively), by: \begin{equation}
I=\frac{2 e}{h} \int dE \, \large[f_R (E,V)-f_L (E,V)\large] \, T (E,V)
, \label{landauer1} \end{equation} where the Fermi filling factors of
the left and right electrodes, $f_L$ and $f_R$ include the bias, and
correspond to the chemical potential of the left and right electrodes
asymptotically inside each electrode.  NDR corresponds to negative
conductance. From Eq.~(\ref{landauer1}), we can derivate with respect
to the bias to obtain the conductance. We obtain two terms, one comes
from the derivative of the Fermi filling factors and the other one
corresponds to the derivative of the transmission.  The derivative of
the Fermi factors is a positive number, hence, only the derivative of
the transmission gives rise to NDR.  It is then, the behavior of the
transmission function with bias that determines the appearance of NDR.

\begin{figure}
\includegraphics[width=0.45\textwidth,clip=true]{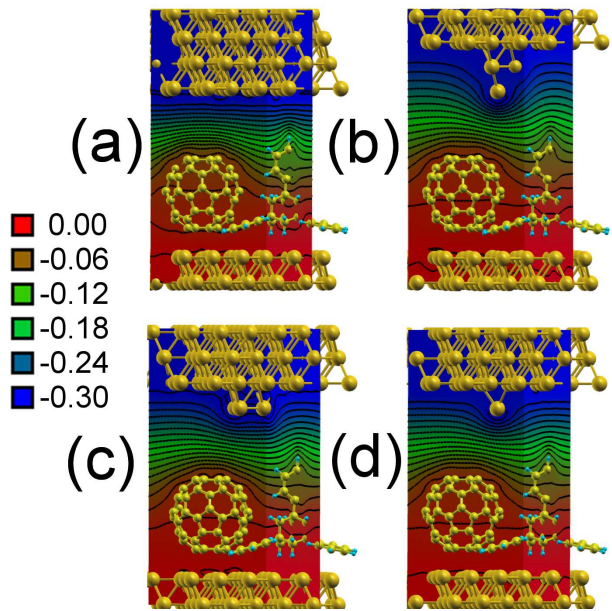}
\caption{Geometry and voltage drop for a 0.3-V bias in the four systems analyzed in this work: (a) a C$_{60}$-TPA overlayer on a semi-infinite
Au(111) an a second semi-infinite
Au(111) representing a
 flat STM tip, the tip-molecule distance is 6.7 \AA ,
(b) the same overlayer under a pyramidal tip formed by 4 atoms and a semi-infinite Au(111) surface, the tip-molecule distance is 6.6 \AA , the tip apex is
off the center of the topmost C$_{60}$ hexagon, 
(c) same as (b) but a capped pyramide with three atoms, the tip-molecule distance is 6.9  \AA , (c) same as (a) plus a gold adatom, the tip-molecule distance is 6.9 \AA , the tip apex is
off the center of the topmost C$_{60}$ hexagon.
The voltage drop is shown on two planes cutting through the C$_{60}$ and the TPA molecule respectively. The plotted isolines are
0.0083 V apart.  
\label{figure1}}
\end{figure}

A quantitative way to compute Eq.~(\ref{landauer1}) is to use
non-equilibrium Green's functions (NEGF) self-consistently solved with
Poisson's equations and fixed chemical potentials in the asymptotic region
inside the electrodes. This has been done using the {\tt Transiesta}
package~\cite{transiesta}, where the electronic structure for the
NEGF is evaluated using density functional theory (DFT) in the local
density approximation (LDA). 
To optimize our calculations, we use a special surface basis-set
that describe the electronic properties at the metal-vacuum
interface~\cite{Sandra}. Hence, the present electronic structure of
the molecular overlayer on Au (111) reproduces the LDA results of
Ref.~\cite{Franke}. The size of the electronic problem is very large
since 5 active layers are used for right electrode, $R$, and four for the left
one, $L$. Each layer contains 28 Au atoms~\cite{Franke}.  The junctions include
one C$_{60}$ molecule and one TPA, plus the STM tip. Figure~\ref{figure1}
shows the geometry and the voltage drop at the STM junctions considered
in this work: $(a)$ a flat tip, $(b)$ a pyramide,
$(c)$ a capped one, and $(d)$ an adatom.
The atom-terminated tips are placed in a low-symmetry
point over the C$_{60}$ molecule, in order to avoid  an unrealistic
high-symmetry configuration.  The two topmost layers of the $R$ electrode
and the topmost one of the $L$ electrode are
relaxed, the other three are fixed to the bulk position, and a recursion
algorithm is used to reproduce semi-infinite electrodes~\cite{transiesta}
while allowing for the correct electric field screening and consequent
Friedel oscillations.

%Figure 1

Figure~\ref{figure1} also shows the voltage drop across the tunneling
junction when the external bias is 0.3 V. We see that most of the bias
drop takes place in the vacuum region. At this bias, the LUMO~\cite{lumo}
of C$_{60}$ is close to the asymptotic chemical potential of the
tip. Hence, C$_{60}$ is largely conducting and the bias drop is 
smaller than in the vacuum gap: the voltage drops a tenth of the total
bias just between the top of the molecule and the metallic substrate. The
molecular states are then subjected to shift under the action of the bias.
However, its partner TPA molecule has a large HOMO-LUMO gap which makes it basically
inert. The voltage drop takes place across the molecule and it amounts
to a seventh of the total bias. Hence, TPA behaves as the dielectric
material of the molecular device that changes the charging energy of the
active molecule,  C$_{60}$. Calculations with and without TPA show that
the charging energy of  C$_{60}$ is reduced by 0.2 eV in the presence
of TPA.  Finally, as the tip becomes sharper, the electric field gets
more localized. As a consequence, the electric field is large close to
the tip and the voltage drop is faster in the vacuum gap while smaller
at the molecular site. Hence, by sharpening the tip we obtain an effect
similar to increasing the vacuum gap which enhances the tunneling-barrier
bias dependence~\cite{Tu}.

\begin{figure}
\includegraphics[width=0.45\textwidth,clip=true]{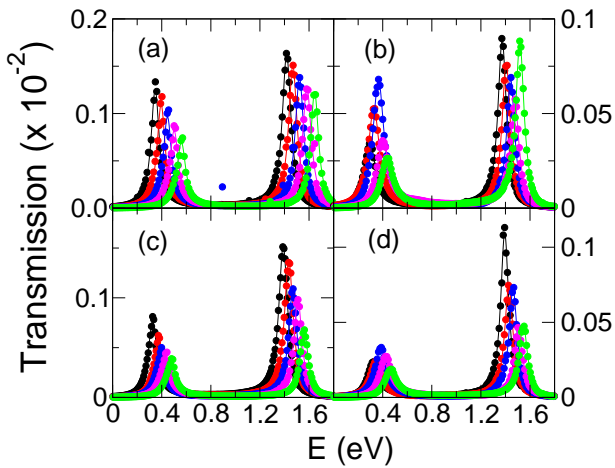}
\caption{Electron transmission as a function of the electron energy (E) 
the same junctions of Fig.~\ref{figure1}: (a) the flat-tip junction (b) 
the pyramidal one (c) the capped one and (d) the adatom one. Filled circles
are {\tt Transiesta} results, and the full line is the fit to
Eq.~(\ref{breit-wigner}). The peaks shift with increasing bias, hence
the lower-energy peak corresponds to 0.0 V, the next one to 0.4 V, 0.8 V,
1.2 V and 1.6 V respectively. Please, notice the different transmission
scales on both sides.  
\label{figure2}} 
\end{figure}

\begin{figure}
\includegraphics[width=0.45\textwidth,clip=true]{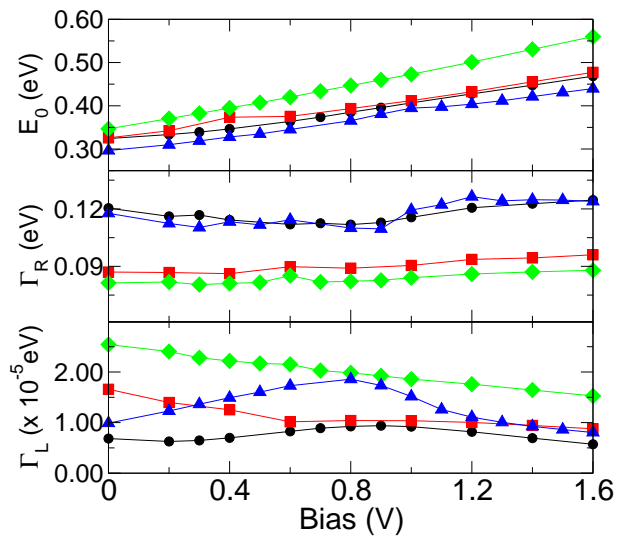}
\caption{
Bias dependence of the LUMO level
$E_0$ with respect to the substrate's Fermi energy, its partial width due to its
coupling to the substrate, $\Gamma_R$
and the partial width due to the coupling to the tip, $\Gamma_L$.
Symbols: diamonds correspond to the flat tip of Fig.~\ref{figure1} $(a)$, 
triangles to the 4-atom tip, $(b)$, 
squares to the capped tip,  $(c)$, and 
dots to the adatom tip, $(d)$.
\label{G}}
\end{figure}

The transmission behavior with  applied bias  varies rapidly as the tip
changes. Figure~\ref{figure2} shows the electron transmission $T$ of
Eq.~(\ref{landauer1}) as a function of the electron energy $E$, as the
bias between tip and surface is increased.  For a fixed bias, $V$, the
transmission function can be reproduced by a Breit-Wigner~\cite{Datta}
formula: \begin{equation} T(E,V)=\frac{\Gamma_R(V)
\Gamma_L(V)}{(E-E_0(V))^2+(\frac{\Gamma_R(V)+\Gamma_L(V)}{2})^2}.
\label{breit-wigner} \end{equation} The full-line curves in
Fig.~\ref{figure2} are fits to Eq.~(\ref{breit-wigner}) and dots are
the {\tt Transiesta}-evaluated transmissions. The agreement is excellent
permitting us to extract the partial widths, $\Gamma_R$ and $\Gamma_L$
as well as the LUMO level $E_0$ as a function of applied bias, see
Fig.~\ref{G}.

The bias dependence of $E_0$, Fig.~\ref{G}, shows that the LUMO level
is not pinned to the substrate's electronic structure because the
bias is not entirely droping in the vacuum gap. This is due to the molecule-molecule
interactions of C$_{60}$ with surrounding TPA, that decouple from the substrate
and polarize C$_{60}$. The flat-tip causes the
largest $E_0$ shift both at zero-bias and at finite bias, showing that
despite the 6.7-\AA~vacuum gap, there is interaction between molecule
and tip.  The tip also modifies the coupling of the molecule with the substrate.
Indeed, the partial width due to the substrate, $\Gamma_R$, can be divided in
two values, Fig.~\ref{G}, one corresponding to the blunt tips $(a)$ and
$(c)$ and the other one to the atomic-like tips $(b)$ and $(c)$.
%The blunt tips have a larger effect on the molecular orbitals, increasing
%their coupling while reducing the substrate's coupling. This can be rationalized
% by a reduction of the substrate-molecule electronic structure hybridization.
Due to  the stronger coupling of the molecule with the Au(111)
substrate, $\Gamma_R$ is larger and has a much smaller $V$-dependence than
$\Gamma_L$ that is the coupling with the tip.  The behavior of $\Gamma_R$
with $V$ can be basically understood by the shift of the LUMO level,
$E_0$. As the bias increases, $E_0$ shifts closer to the vacuum level of
the substrate, slightly increasing the partial width $\Gamma_R$.  However,
$\Gamma_L$ is strongly dependent on the actual density of states (DOS)
of the tip's electronic structure.  Sharp tips, $(b)$ and $(d)$, show a
non-monotonic behavior of $\Gamma_L$ with $V$. This is also seen
in the electron transmission since by virtue of Eq.~(\ref{breit-wigner}),
the transmission at $E_0$ is roughly proportional to $\Gamma_L$.

\begin{figure}
\includegraphics[width=0.40\textwidth,clip=true]{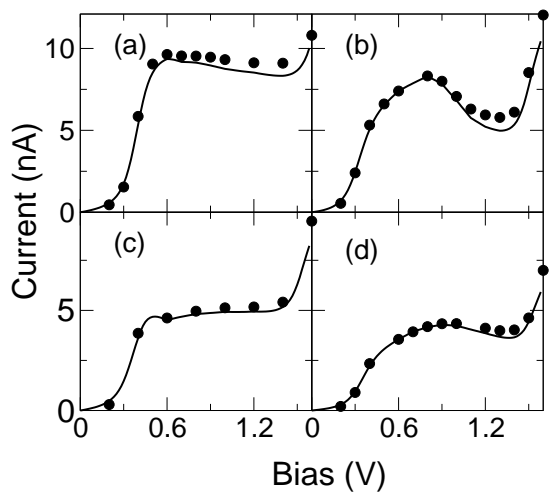}
\caption{I-V characteristics for the four cases studied here in the same order as Fig.~\ref{figure1}. Dots are the {\tt Transiesta} results, and full line the 
results of Eq.~(\ref{landauer1}) using the Breit-Wigner fit,
Eq.~(\ref{breit-wigner}), interpolating the LUMO and LUMO+1 levels and
partial widths for bias in between the {\tt Transiesta} calculations.
The fit excellently reproduces the LUMO onset in the current, however
errors accumulate in between the LUMO and LUMO+1 onsets where the
Breit-Wigner fit is worse.  The current peak-to-valley ratios (dots)
are $(a)$ 1.07, $(b)$ 1.4, $(c)$ -- , $(d)$ 1.1, the experimental
one~\cite{Franke} is 1.3.  
\label{figure3}} 
\end{figure}

 Figure~\ref{figure2} $(a)$ is the transmission function for the
flat tip case. We see that the transmission maximum, corresponding
to the LUMO, shifts to higher energies as the bias is increased. This
is due to the partial bias drop at the center of the molecule and is
reproducing the behavior described in Ref.~\cite{Tu}. More interestingly,
the maximum height drops with bias. This is the behavior described
in Ref.~\cite{Grobis} causing NDR. And indeed, as can be seen in
Fig.~\ref{figure3} $(a)$ NDR is found for this system. This fast decay
of the transmission is due to the increase of the tunneling barrier with
increasing bias~\cite{Grobis,Tu}.

%Figure 2

For the 4-atom pyramide tip, Fig.~\ref{figure2} $(b)$, the maximum of
the transmission peaks when the bias is at $\sim$0.8 V. This is totally
different from the behavior of Fig.~\ref{figure2} $(a)$ and the one
described in  Ref.~\cite{Grobis}. The maximum of the transmission with
bias can be traced back to a resonance of the 4-atom pyramidal tip of
atomic origin. This is found at -0.45 eV in the DOS of the tip. Then,
the strong variation of $\Gamma_L$ with bias, Fig~\ref{G}, is due to
the resonance in the DOS. When the tip is capped, Fig.~\ref{figure1}
$(c)$, the corresponding $\Gamma_L$ recovers a monotonic decrease
with bias as for the flat-tip case $(a)$, Fig.~\ref{G}. 
Hence, the transmission is very similar to the case
of the flat-tip. Finally, case $(d)$ where one adatom is added to the
flat tip shows the same kind of non-monotonic behavior in $\Gamma_L$
as in $(b)$ and a clear maximum in both, $\Gamma_L$ and the transmission peaks for the LUMO
around 0.8 V.

%Figure 2bis

Figure~\ref{figure3} presents the $I$--$V$ characteristics for the
four systems of Fig.~\ref{figure1}.  The NDR features are large for the
four-atom pyramidal tip, Fig.~\ref{figure3} $(b)$ and  small for the flat
tip, Fig.~\ref{figure3} $(a)$. Hence, the largest NDR is obtained when
the partial widths present a resonant behavior with the applied bias,
as the one due to electronic resonances in the electrodes, tip $(b)$. The
monotonic bias dependence of partial widths due to the barrier dependence
on the bias, case $(a)$, is slower leading to weak NDR.  It is interesting
to study the capped pyramidal tip,  Fig.~\ref{figure3} $(c)$, where NDR
is absent and the $I$--$V$ curve is closer to the flat-tip one. Here,
the partial width dependence is slow, Fig.~\ref{G}, with increasing
bias because the tip's DOS is smoothly increasing. Despite the fact
that the barrier height increases with bias, reducing the partial width,
the increase of DOS is large enough to partially compensate the larger
barrier and the transmission only drops slowly, Fig.~\ref{figure2}
$(c)$. NDR disappears. However, if one adatom is added to the flat tip,
a bigger NDR effect is obtained because the sharper tip DOS forces a
resonant-like $\Gamma_L$ and hence  a peak-shaped transmission.
It is then this last effect, rather than the enhancing of the bias dependence
of the tunneling barrier, Fig.~\ref{figure1}, that drives NDR in sharp tips.

The shape of the $I$--$V$ characteristics changes with tip. While the flat
tip has a sharp onset and a sharp peak at 0.6 V, the atom-terminated tips
have a broad peak centered at 0.8 - 0.9 V with a $\sim$ 0.5-V width. The
experimental one is a broad peak centered at 1.1 V with a $\sim$ 0.5-V
width~\cite{Franke}. This fact and the current peak-to-valley ratio,
Fig.~\ref{figure3}, render the pyramidal tip  in very good agreement with
the experiment. Moreover, the experimental distance between the LUMO and
LUMO+1 onsets~\cite{Franke} is 1.5 V and the one in Fig.~\ref{figure3}
is 1.2 V. The actual $I$--$V$ shape is then determined by the bias dependence
of the level, $E_0$, and of the partial widths, here $\Gamma_L$.
Indeed, the shape of, for instance, the LUMO onset is given by the
$V$-dependence of $E_0$~\cite{Gauyacq}, while the NDR drop is dominated
by the $V$-dependence of $\Gamma_L$.

%Figure 3

In summary, we have performed NEGF calculations 
for the experimental setup of Ref.~\cite{Franke}. 
Our simulations have permitted us to rationalize the appearance of NDR
in this system in terms of molecular levels and partial widths because
 we have been
able to extract their bias dependence. We have shown that NDR can be
tuned by modifying molecule-molecule and molecule-electrode interactions.
The molecule-molecule interactions induce a reduction of the overwhelming
presence of the substrate permitting us to have molecular properties
with sizeable bias dependence. On the other hand, molecule-electrode
interactions determine the molecular partial widths. NDR effects are
maximum when the partial widths vary rapidly with the applied
bias. This dependence is largest in the presence of electronic resonances
on the electrodes such as the ones obtained for sharp STM tips. Finally, our
results show that the overall shape of the $I$--$V$ characteristic contains
relevant information on the molecule-molecule and molecule-electrode
interactions.

\begin{acknowledgments}
We thank  K. J. Franke and J. I. Pascual for very interesting discussions.
F.D.N acknowledges support from Juan de la Cierva program.  Computing
resources from CESGA and financial support from the japanish JSPS,
the spanish MICINN (FIS2009-12721-C04-01), and the european ICT project
``AtMol'' are gratefully acknowledeged.  
\end{acknowledgments}


\begin{thebibliography}{99}
\bibitem{lumo} The LUMO of C$_{60}$ is three-fold degenerate but there is
only one LUMO contributing to the largest transmission eigenchannel.
\end{thebibliography}
\end{document}